\documentclass[proof]{pasj00}

\usepackage{color}
\usepackage{booktabs}
\usepackage{lscape}
\usepackage{mathptmx}
\usepackage{subfigure}

\begin{document}
\SetRunningHead{
Repeated Short-term Spectral Softening in the Low/Hard State of Swift J1753.5-0127
}{
A. Yoshikawa et al.
}
\title{
Repeated Short-term Spectral Softening in the Low/Hard State of
the Galactic Black-Hole Candidate Swift J1753.5-0127 \\
}

\author{
Akifumi \textsc{Yoshikawa}\altaffilmark{1,2}, Shin'ya \textsc{Yamada}\altaffilmark{1}, Satoshi \textsc{Nakahira}\altaffilmark{3}, Masaru \textsc{Matsuoka}\altaffilmark{4}, Hitoshi \textsc{Negoro}\altaffilmark{5}, Tatehiro \textsc{Mihara}\altaffilmark{4}, and Toru \textsc{Tamagawa}\altaffilmark{1}
}
\altaffiltext{1}{
   High Energy Astrophysics Laboratory, RIKEN Nishina Center\\
   2-1 Hirosawa, Wako-shi, Saitama, 351-0198, Japan
   }
\email{akifumi@crab.riken.jp}
\altaffiltext{2}{
   Department of Physics, Tokyo University of Science, 1-3 Kagurazaka, Shinjuku-ku, Tokyo 162-8602, Japan
   }
\altaffiltext{3}{
   ISS Science Project Office, ISAS, JAXA, 2-1-1 Sengen, Tsukuba, Ibaraki 305-8505, Japan
   }
\altaffiltext{4}{
   MAXI team, RIKEN, 2-1, Hirosawa, Wako, Saitama 351-0198, Japan
   }   
\altaffiltext{5}{
   Department of Physics, Nihon University, 1-8-14 Kanda-Surugadai, Chiyoda-ku, Tokyo 101-8308, Japan
   }   
   

\Received{$\langle$Jun. 30th 2013$\rangle$}
\Accepted{$\langle$accepted date$\rangle$}
\Published{$\langle$publication date$\rangle$}
\maketitle

\begin{abstract} 
We report MAXI and Swift observations of short-term spectral softenings of the galactic black-hole X-ray binary Swift J1753.5-0127 in the low/hard state.
These softening events are characterized by a simultaneous increase of soft X-rays (2--4~keV) and a decrease of hard X-rays (15--50~keV) lasting for a few tens of days.
The X-ray energy spectra during the softening periods can be reproduced with a model consisting of a multi-color disk blackbody and its Comptonized component.
The fraction of the Comptonized component decreased from 0.30 to 0.15 when the spectrum became softer; 
meanwhile the inner disk temperature ($T{_{\rm{in}}}$) increased from 0.2 to 0.45~keV.
These results imply that the softening events are triggered by a short-term increase of the mass accretion rate.
During the observed spectral softening events, the disk flux ($F_{\rm{disk}}$) and $T{_{\rm{in}}}$ did not obey the $F{_{\rm{disk}}} \propto T{_{\rm{in}}}$~${^{4}}$ relation,
suggesting that the inner disk radius does not reach the innermost stable circular orbit.
\end{abstract}

\section{Introduction}
X-ray emission from black-hole binaries (BHBs) comes from the release of the gravitational energy of accreting matter falling into a black hole. 
Mass accretion onto the black hole in low-mass X-ray binaries (LMXBs) proceeds via Roche lobe overflow from a relatively light companion star. 
According to X-ray observations over the last four decades, BHBs which are mostly observed in LMXBs, appear to be transient \citep{RandC06}. 
There are a few LMXB sources that are observed to be persistent (e.g., 1E 1740.7-2942: \cite{Cui97} and GRS 1758-258: \cite{Cui02}).
However, their properties have not yet been studied in details, due to a lack of their observations. 
BHBs in LMXBs exhibit relatively short timescale outbursts ($\sim$weeks or months) and long timescale quiescence ($\sim$years). 
Both timing and spectral properties change vastly as the outburst evolves. 
Specifically, there are three representative states: 
the low-luminosity hard (low/hard) state, the high-luminosity soft (high/soft) states \citep{RandC06} and the intermediate state which appears during transition from the high/soft to the low/hard state. 

X-ray spectra in the low/hard state can be well reproduced by a powerlaw with a photon index of 1.4--2.1 \citep{RandC06} and a high-energy cut-off at $\sim$100~keV, which is interpreted as the result of inverse-Compton X-ray emission from the hot plasma around the black hole.
In the high/soft state, the time variability is quite suppressed, 
and the energy spectrum is soft.
The spectrum can be described well by the optically thick thermal disk emission (cf., disk black body ({\bf diskbb} model): \cite{Mit84}) accompanied by a steep powerlaw with a photon index of 2--2.5 \citep{Don07}.
A conceivable picture of this state is an optically thick and geometrically thin disk, the so-called the standard disk \citep{SandS73}, which extends down to the inner most stable circular orbit (ISCO).
This picture is supported by many studies (e.g., \cite{Max86} and \cite{Ebi91}).
Transitions between the low/hard and high/soft state during the outburst have been observed in many BHBs, 
and the energetic interaction between the disk and the corona during the state transitions
 has been modeled as a functions of time and the mass accretion rate \citep{RandC06}.

\begin{figure*}
   \begin{center}
      \FigureFile(180mm,150mm){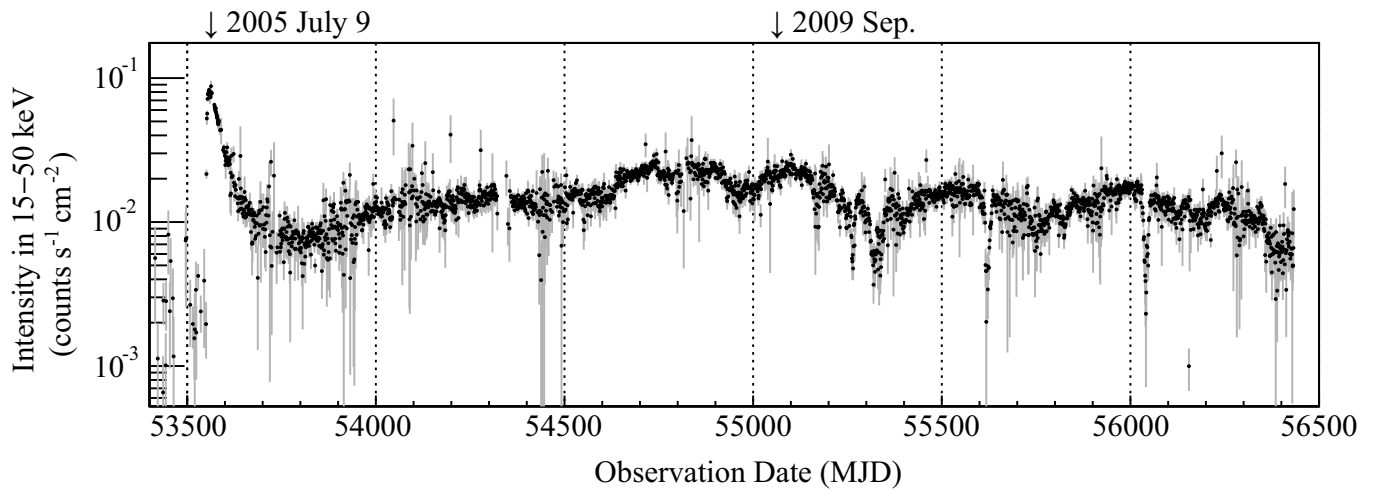}
   \end{center}
   \caption{
     The light curve in the 15--50 keV energy band observed by the BAT.
     On 2005 July 9, Swift J1753.5-0127 was discovered by the detection of the outburst.
     In 2009 Sep., MAXI started operating on board the International Space Station.
   }\label{FIG:LC1550}
\end{figure*}

Swift J1753.5-0127 is an LMXB discovered with the Burst Alert Telescope (BAT) on board ${\it Swift}$ on 2005 May 30 (\cite{atel546} and \cite{atel547}).
The light curve in the 15-50 keV energy band is shown in figure \ref{FIG:LC1550}.
The value of the inner disk temperature ($T_{\rm{in}}$) and the photon index during the outburst of 2005 was estimated to be 0.2--0.4~keV and 1.6--1.8 (\cite{Ra07}, \cite{Mil06} and \cite{Sa08}).
Its binary orbital period is reported to be 3.23~hrs \citep{Zu08}, which is the third shortest period of all the BHBs thus far.
The distance to the source and an inclination angle of the disk have not been precisely determined \citep{Bel07}.
After the discovery, the source flux peaked at $\sim$200~mCrab on 2005 July 9, and then gradually decreased for more than 6~months.
Since 2006, the flux gradually increased again, becoming as bright as $\sim$20~mCrab.
Such a long lasting outburst is highly unusual in LMXBs.
Dips in the 15--50 keV energy range (hard-dips) lasting for $\sim$25~days were observed by the BAT.
Dips in the 2--20~keV energy range, which are interpreted as a possible eclipse by the warped disk, have also been reported \citep{SH13}.
However, the state transition and quiescence of Swift J1753.5-0127 has not yet been observed \citep{Ra07}.
Continuous monitoring of Swift J1753.5-0127 by MAXI first revealed a short-term spectral softening lasting for a few tens of days in 2009 (\cite{atel2341} and \cite{Ne10}). 
When the source showed another spectral softening in 2012, we proposed target of opportunity observations (ToOs) using the Swift X-ray Telescope (XRT), and succeeded in tracing its spectral time evolution. 

In this paper, we report that the spectral and temporal properties obtained during the spectral softenings of Swift J1753.5-0127 and suggest that the softenings are caused by the state transitions.
Our observations and data reduction are presented in section 2.
In section 3, the spectral analysis is presented. 
We discuss these results in section 4 and summarize them in section 5.

\section{Observation and Data Reduction \label{SEC:2}}
\subsection{Data reduction of MAXI and Swift \label{SEC:DATA}}
In 2009 September, the Monitor of All-sky X-ray Image (MAXI: \cite{Matu09}) started operating on board the International Space Station (ISS). 
The Gas Slit Camera (GSC: \cite{Miha11} and \cite{Sugi11}) and the Solid-state Slit Camera (SSC: \cite{Tsu10} and \cite{Tom11}) on board MAXI have energy ranges of 2--20~keV and 0.7--7.0~keV, respectively. 
The GSC is composed of 12 position sensitive proportional counters with six carbon anodes in each.
The SSC is a slit camera with an array of 32 X-ray CCD chips, each of which is one inch squares.
We extracted the GSC data from the event files processed with the MAXI standard analysis software (\cite{Sugi11} and \cite{Naka13} \footnote[1]{http://repository.tksc.jaxa.jp/pl/dr/AA0061914004}).
The event files include the data obtained by the GSC operated with the standard operation voltage of 1650~V and a reduced voltage of 1550~V.
We extracted the source photons from a circle centered at the target position with a radius of 2$^\circ$.0 and the background photons from a annular region with an inner and an outer radius within 2$^\circ$.0 and 3$^\circ$.0.

Because of the various background components in the SSC data \citep{Tsu10},
we processed the SSC data from the event files in the following order.
We first discarded the hot pixel events using the {\it cleansis} ftool.
To exclude photons from the sun or the moon, 
we accumulated events over a good time interval \citep{Dai10}.
We used the GRADE0 \citep{Tom11} events below 505~ch ($\sim$1.85~keV = Si edge) and the GRADE012 events larger than 505~ch.
We extracted the source photons from a circle centered at the target position with a radius of 1$^\circ$.5, and the background photons from a circle with a radius of 3$^\circ$.0 excluding a region within 1$^\circ$.5 from the target. 

Swift performed the ten ToOs of Swift J1753.5-0127 on MJD 56037, 56039, 56041, 56043, 56045, 56047 56049, 56050, 56053 and 56055 (corresponding to dates from 2012 April 20 to May 8).
These observations were all performed in the Windowed Timing (WT) mode to avoid pile up events.
The net exposures were 1.9, 1.5, 1.0, 1.2, 0.5, 1.0, 0.8, 1.1, 1.0, and 1.1~ksec, respectively.
We reduced the XRT archival data using the web tools \footnote[2]{http://www.swift.ac.uk/user$\_$objects/} which was provided by the UK Swift Science Data Center at the University of Leicester.

\subsection{Light curves and hardness ratios \label{SEC:MS}}
\begin{figure*}
   \begin{center}
      \FigureFile(120mm,100mm){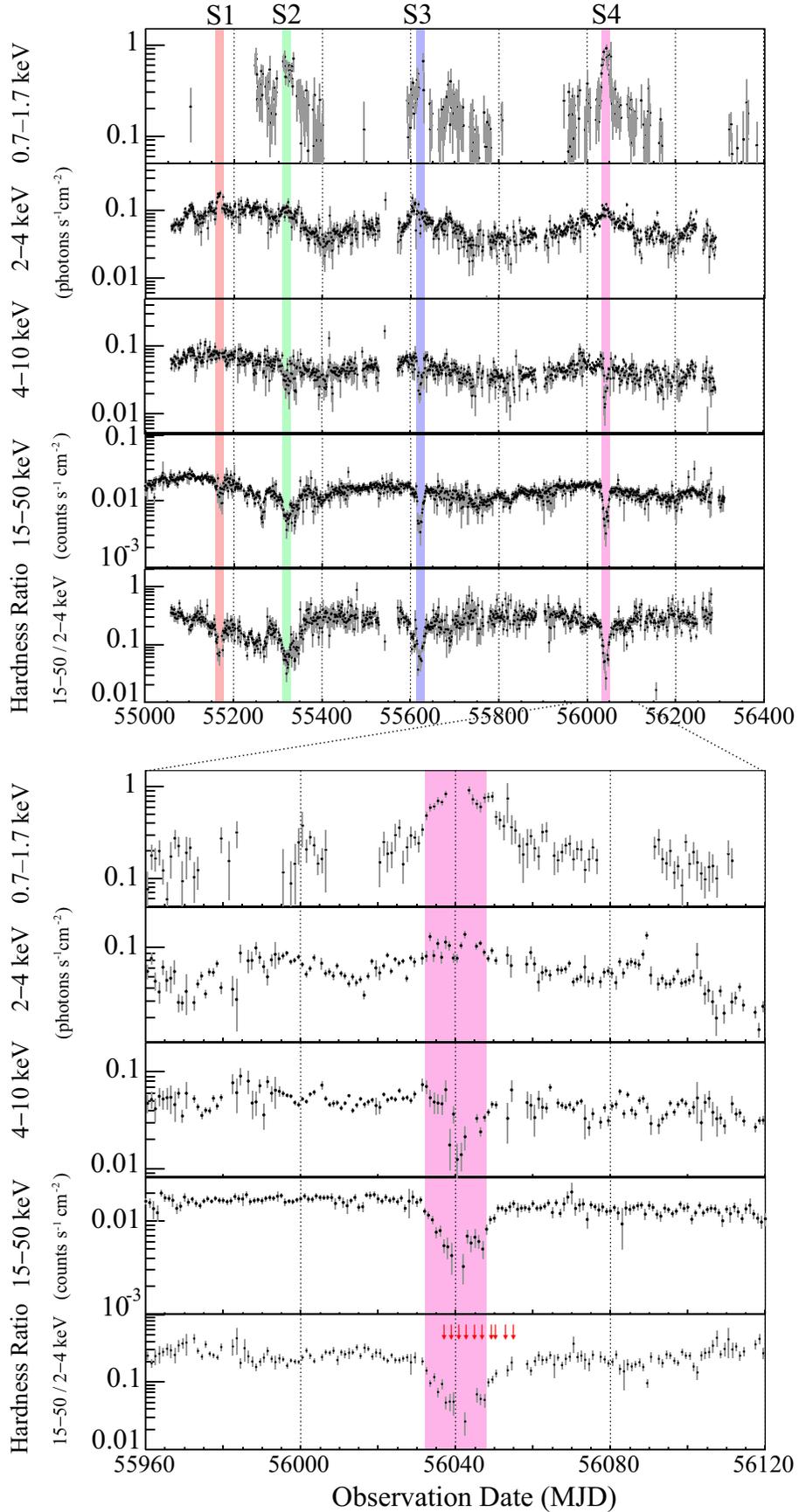}
   \end{center}
   \caption{
          (Upper panel) The light curves in the energy band of 0.7--1.7~keV (top), 2--4~keV (second), and 4--10~keV (third), and 15--50~keV (fourth) and the hardness ratio of 15--50~keV to 2--4~keV (bottom) from MJD 55000 to 56400.
          The light curves in the energy bands of 0.7--1.7~keV, 2--10~keV, and 15--50~keV were observed by the SSC, GSC and BAT, respectively.
          Hatched regions in red, green, blue and magenta are S1, S2, S3 and S4, respectively.
          (Lower panel) The light curves from MJD 55960 to 56120.
          Red arrows show the XRT observations in S4.
   }\label{FIG:LC}
\end{figure*}

In figure \ref{FIG:LC}, we show the light curves in the energy ranges of 0.7--1.7~keV (SSC), 2--4~keV, 4--10~keV (GSC), and 15--50~keV (BAT) and the hardness ratio of 15--50~keV to 2--4~keV.
Each time bin of the light curves is a day, which is long enough to obtain high signal to noise ratios. 
There are gaps in the GSC light curves between MJD 55529 and 55570, 
because the source was out of the FOV of the GSC 
during the period the source was close to the direction of the rotation axis of the ISS. 
SSC data were fewer than GSC data, because the FOV of the SSC was narrower than that of the GSC.

\begin{figure*}
   \begin{center}
      \FigureFile(180mm,160mm){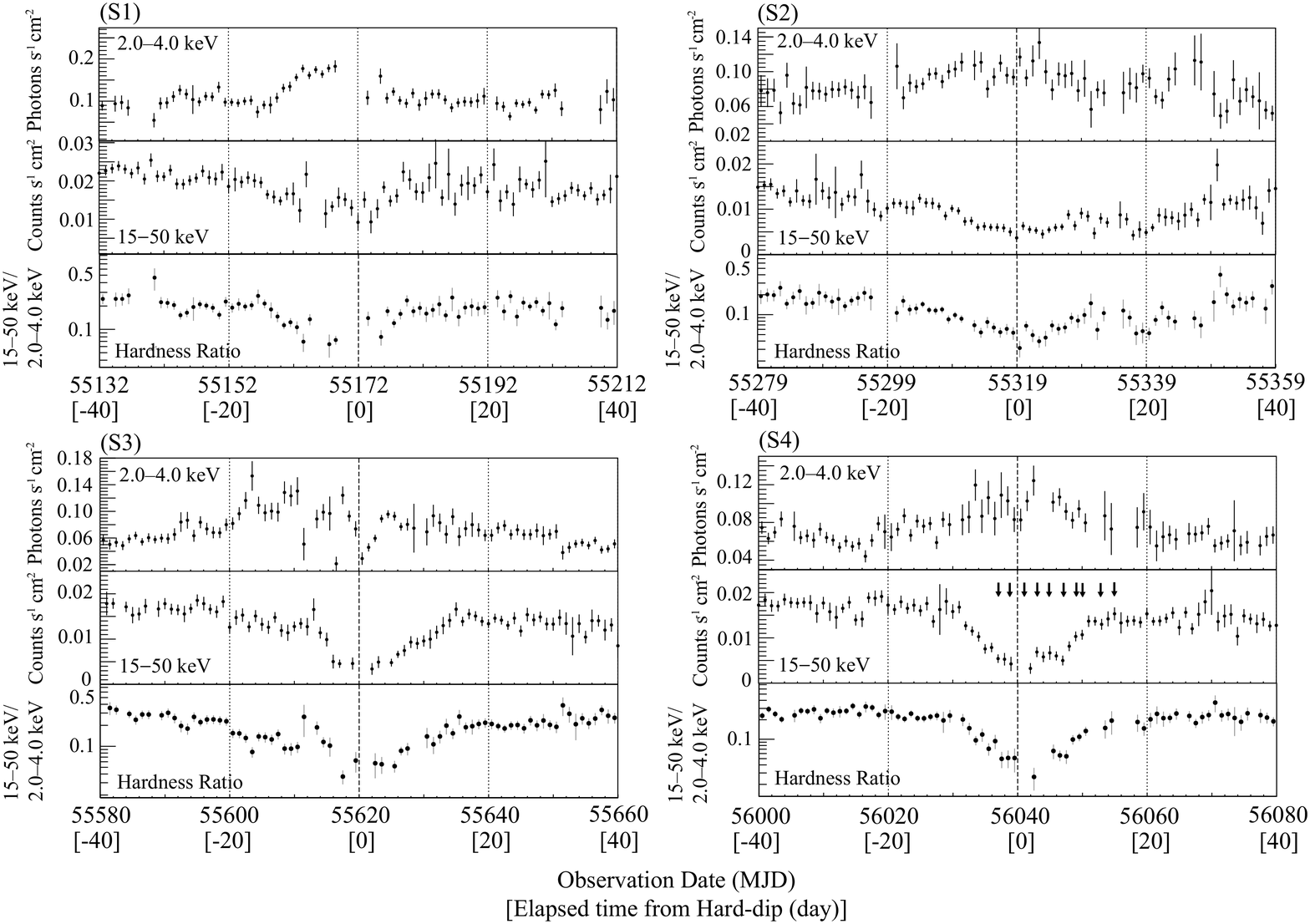}
   \end{center}
   \caption{
          (top left) Light curves in the energy bands of 2--4~keV (top), and 15--50~keV (middle), and the hardness ratio of 15--50~keV to 2--4~keV (bottom) during S1.
          The light curves and hardness ratio during S2 (top right), S3 (bottom left) and S4 (bottom right). 
          Origins correspond to the local minimum of the light curves in the energy range of 15--50~keV observed by the BAT.
          Arrows show the XRT observations in S4.
   }\label{FIG:4Ses}
\end{figure*}

The four short-term softenings on MJD 55172 (2009 Dec. 7), 55319 (2010 May 3), 55620 (2011 Feb. 28) and 56040 (2012 Apr. 23) were characterized by an increase of soft X-rays (2--4~keV) and a decrease of the hard X-rays (15--50~keV) lasting for a few tens of days, as shown in the lower panel of figure \ref{FIG:LC}.
Hereafter, we call the four events as S1 (MJD 55164--55180), S2 (MJD 55311--55327), S3 (MJD 55612--55628), and S4 (MJD 56032--56048), respectively. 
The time scales of each softening were below 40~days, as shown in figure \ref{FIG:4Ses}.

\begin{figure}
   \begin{center}
      \FigureFile(90mm,60mm){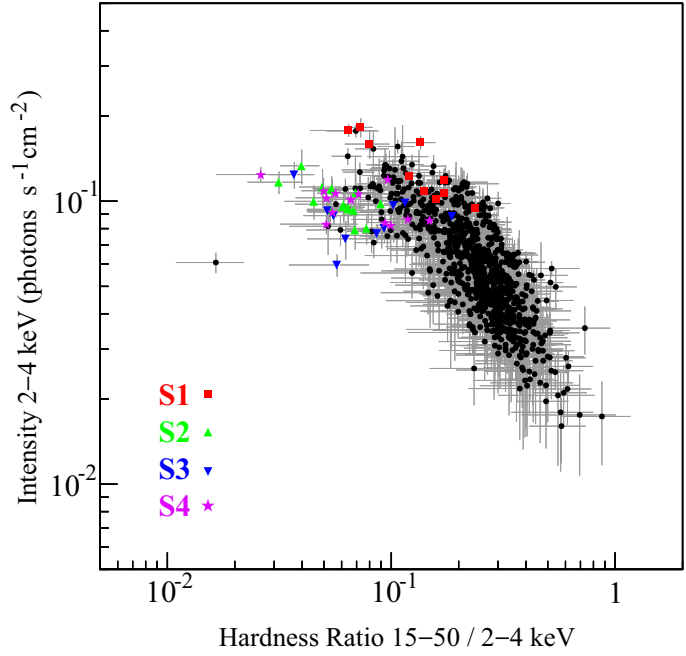}
   \end{center}
   \caption{
	X-ray hardness intensity diagram of Swift J1753.5-0127 since the launch of MAXI. 
        Hardness refers to the ratio of 15--50~keV to 2--4~keV count rates
        and intensity means the 2--4~keV count rate.
	Data in S1, S2, S3, and S4 are shown in red, green, blue, and magenta, respectively.
   }\label{FIG:HR}
\end{figure}

Figure \ref{FIG:HR} is a hardness intensity diagram (HID) showing the relation between the intensity in the 2--4~keV band and the hardness ratio of 15--50~keV to 2--4~keV.
When the intensity in the 2--4~keV energy band increased, the X-ray spectra became softer.
The data points for S1, S2, S3 and S4 all lie in the region where the hardness ratio is lower and the 2--4~keV intensity is higher.
The HID between the hardness ratio (6--10~keV / 3--6~keV) and the intensity (2--15~keV) based on RXTE/PCA during and after the outburst of 2005 (MJD 53500--55500) has been plotted by \citet{Sol13}.
Their HID shows the failed transition into the high/soft state during the outburst.
The data points for the failed transition are softer than those of S1, S2, S3, and S4 in figure \ref{FIG:HR}.

\section{Spectral Analysis \label{SEC:3} } 
\begin{figure}
   \begin{center}
      \FigureFile(90mm,60mm){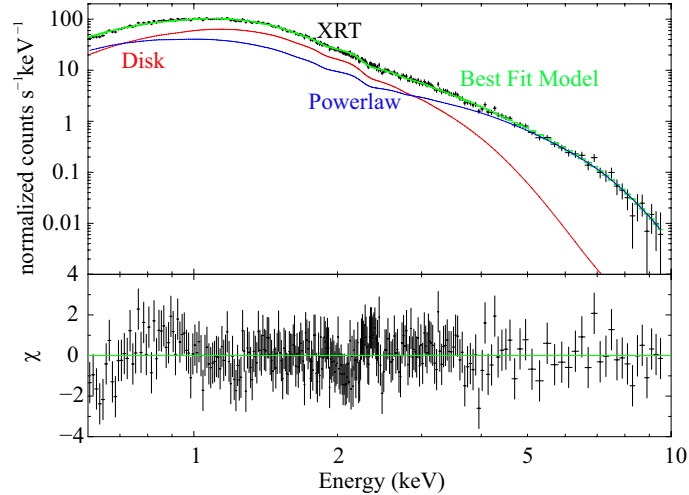}
   \end{center}
   \caption{
	The spectrum of XRT on MJD 56041 is shown in black.
        The best-fit {\bf tbabs*(diskbb+powerlaw)} model,
        the {\bf diskbb} component, and the {\bf powerlaw} component are shown as green, red, and blue lines, respectively.
	Residuals between the data and best-fit model are shown in the lower panel.
   }\label{FIG:XRTCntsSpe}
\end{figure}

We performed spectral analyses of the XRT data in S4, because the source was more frequently observed in S4 with XRT than in the other periods.
An example of the XRT spectra is shown in figure \ref{FIG:XRTCntsSpe} based on the observation of MJD 56041.
The spectral fit was performed with XSPEC version 12.7.1.
The spectral fitting was performed from 0.6 to 9.4 keV, the response file (swxwt0to2s6$\_$20110101v015.rmf) was designated, and the arf files were generated in the web tools.
A systematic error of 3$\%$ was taken into account, given the uncertainties of the response matrix \footnote[3]{http://www.swift.ac.uk/analysis/xrt/spectra.php}.

We fitted all the data with the Comptonized disk emission.
We applied an empirical {\bf diskbb} plus {\bf powerlaw} model,
for the optically disk emission and the Comptonized emission, respectively.
We adopted the {\bf tbabs} model and the {\bf wilm} abundance table for the interstellar absorption.
To estimate $N_{\rm{H}}$ we summed four XRT spectra on MJD 56037, 56039, 56041, and 56043, in which significant {\bf diskbb} components and absorption were present, and fitted a model to the summed spectrum.
The model is written as a {\bf tbabs *(diskbb + powerlaw)} in the XSPEC terminology.
The estimated value of $N_{\rm{H}}$ was 2.8 $\times$ 10${^{21}}$~cm$^{-2}$, which is close to the value obtained in the previous work (\cite{Re09}, \cite{Bel07} and \cite{Sol13}).
Fixing $N_{\rm{H}}$ to 2.8 $\times$ 10${^{21}}$~cm$^{-2}$, we then fitted the {\bf tbabs *(diskbb + powerlaw)} model to all the XRT spectra.
The best-fit models and the residuals are shown in figure \ref{FIG:XRTCntsSpe} based on the observation of MJD 56041.
The best-fit parameters are listed in table \ref{table:PL}. 
As can be seen from table \ref{table:PL} and figure \ref{FIG:XRTCntsSpe}, the model reproduced the data well.
A powerlaw with a photon index of $\sim$2.1--2.9 and dominant soft component characterized the high/soft state of BHBs.
The disk temperature gradually decreased during our XRT observations.

\begin{figure*}
   \begin{center}
      \FigureFile(180mm,100mm){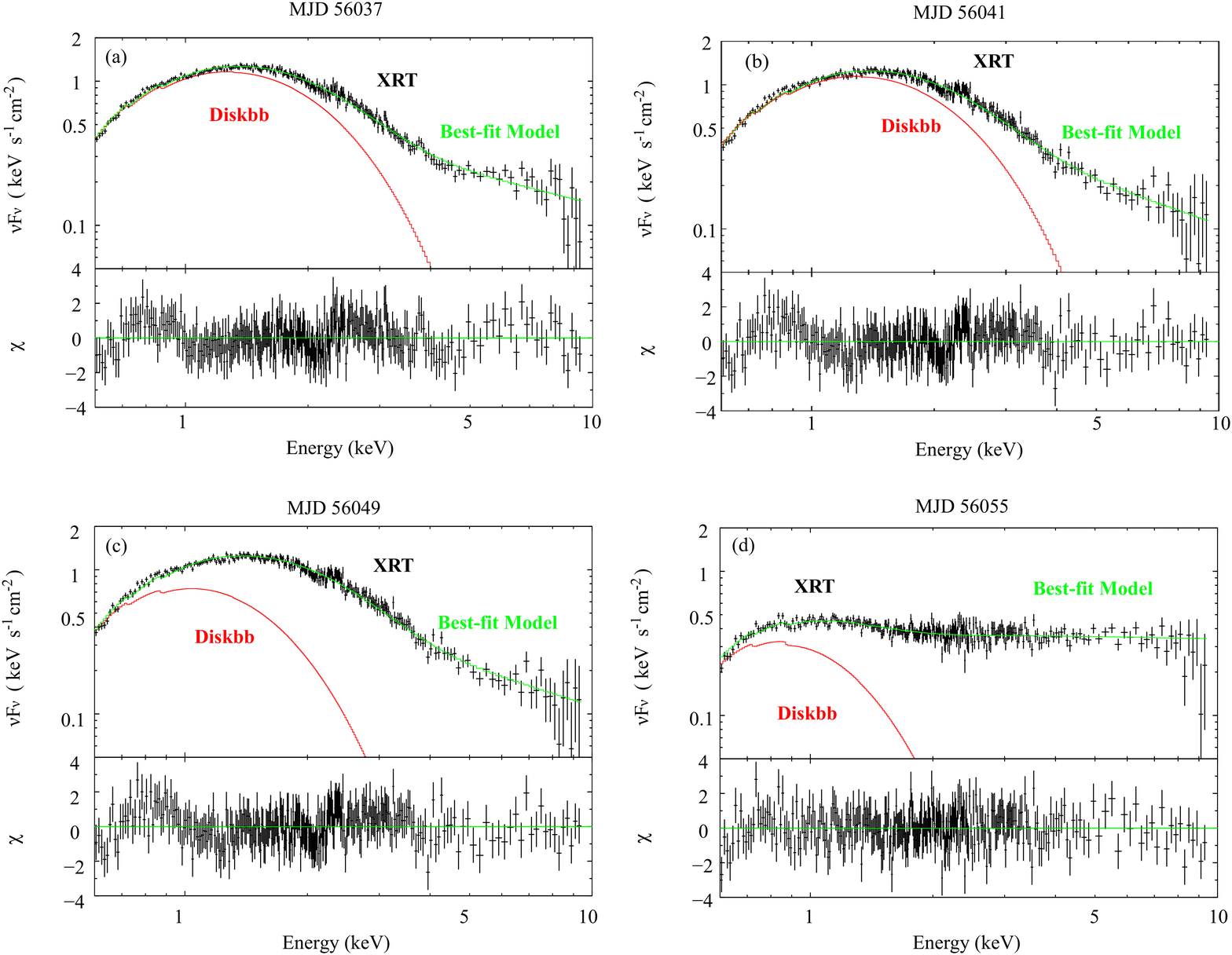}
   \end{center}
   \caption{
    Time-averaged spectra of Swift J1753.5-0127 unfolded with the response in ${\nu}F_{\nu}$ form, extracted from the data of MJD 56037 (top left), 56041 (top right), 56049 (bottom left), and 56055 (bottom right) observed by Swift XRT.
    The best-fit {\bf tbabs*simpl*diskbb} model (green) and the {\bf tbabs*diskbb} component (red) are overlaid.
    Residuals between the data and best-fit model are shown in the lower panels.
   }\label{FIG:EFE}
\end{figure*}

Second, in order to discuss the {\bf diskbb} flux precisely, we replace the {\bf powerlaw} model with the {\bf simpl} model \citep{Ste09}.
The {\bf simpl} model is an phenomenological convolution model that converts a given fraction of the incident spectrum into a powerlaw for the Comptonized emission.   
This model is written as a {\bf tbabs*simpl*diskbb} in the XSPEC terminology.
It was difficult to constrain the photon index ($\Gamma$) of the {\bf powerlaw} model without any information above $\sim$10~keV.
Therefore, we constrained $\Gamma$ to lie within 1.8--3.0, which is reasonable for the hard tail of BHBs in the high/soft state \citep{RandC06}.
We also obtained acceptable fits with this model for all the spectra.
To fit the {\bf tbabs*simpl*diskbb} model to the spectra, the value of $N_{\rm{H}}$ was estimated to be 2.0 $\times$ 10${^{21}}$~cm$^{-2}$ in a similar way to the fitting the {\bf tbabs *(diskbb + powerlaw)} model.
The value is also similar to the value obtained in the previous work \citep{Rey10}.
Since {\bf simpl} redistributes input photons to higher and lower energies,
 the energy range of the response matrix was extended from 0.1 to 1000 keV.
Figure \ref{FIG:EFE}a--d shows four examples of the XRT spectra that were observed from the beginning to the end of the softening event.
The best-fit {\bf tbabs*simpl*diskbb} models are superposed and the residuals are shown in the lower panel of each figure.
The best-fit parameters are listed in table \ref{table:Sim}.
The fitting result in MJD 56039 was not acceptable.
When we fit the spectrum in MJD 56039 with the model leaving $N_{\rm{H}}$ free, 
the fitting result was improved and the value of $N_{\rm{H}}$ decreased.
This probably implies a short-term variation of $N_{\rm{H}}$. 
To avoid the uncertainty, we ignored the data below 1.0 keV in the fitting.
We will mainly discuss the best-fit parameters of the {\bf tbabs*simpl*diskbb} model, because the {\bf simpl} model can take into account the Comptonized photons more properly than the {\bf powerlaw} model \citep{Ste09-02}.

\begin{figure}
   \begin{center}
      \FigureFile(90mm,60mm){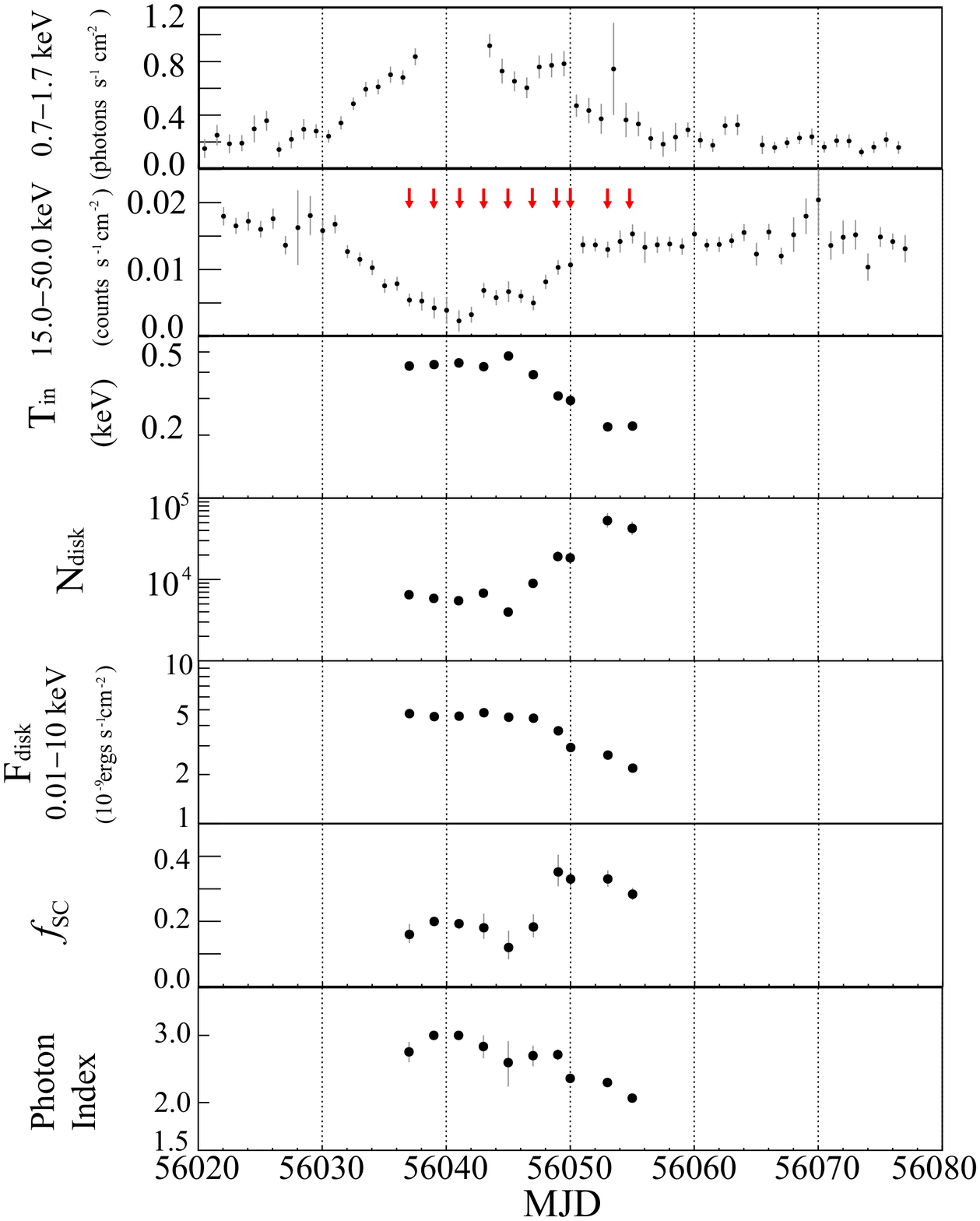}
   \end{center}
   \caption{
	  Time evolution of the best-fit parameters of the {\bf tbabs*simpl*diskbb} model and X-ray intensity. 
	  From top to bottom, the intensity in 0.7--1.7~keV observed by the SSC, that in 15--50~keV observed by the BAT, ${T_{\rm{in}}}$, $N_{\rm{disk}}$, $F_{\rm{disk}}$, the scattered fraction ($f_{\rm{sc}}$), and photon index.
          The parameter $f_{\rm{sc}}$ indicates the rate of conversion of photons from black-body emission into Comptonized photons.
          Some error bars are hard to see, because they are smaller than the plotting symbols.
          Red arrows show ten XRT observations.
   }\label{FIG:LCPa}
\end{figure}

Figure \ref{FIG:LCPa} shows the time evolution of the fitting results in S4.
The best-fit parameters appeared to reach a plateau between MJD 56037 and 56043.
The best fit value of $T_{\rm in}$ gradually decreased from $\sim$0.4~keV to $\sim$0.2~keV from MJD 56047 to MJD 56055.
The lowest value of $T_{\rm in}$ is close to that obtained by Suzaku observation of the low/hard state in 2007 \citep{Rey10}.
However, a powerlaw with a photon index of 2.06 on MJD 56055 still characterized the high/soft state.
These results suggest that the source is in the intermediate state.
The values of $T_{\rm{in}}$ and the disk flux in the 0.01--10~keV energy range ($F_{\rm{disk}}$) 
changed from 0.48 to 0.22~keV and from 4.5 $\times$10$^{-9}$ to 2.2 $\times$10$^{-9}$~erg~s$^{-1}$~cm$^{-2}$, respectively.
$F_{\rm{disk}}$ declined and $f_{\rm{sc}}$ increased from MJD 56045 to 56055, 
which indicates a state transition from the soft to hard state, as shown in figure \ref{FIG:LCPa}.

To examine an existence of Fe K$\alpha$ emission due to disk reflection,
we added a broad gaussian to the model for the data based on MJD 56041, but the fits were not significantly improved,
which was consistent with there being little enhancement in the raw spectra and few observations around 6~keV (see figure \ref{FIG:XRTCntsSpe}).
Thus, we could not confirm the existence of the broad iron lines, which would be presumably due to the fitting methods: fitting the spectra ignoring the 4--7~keV band, for instance \citep{Re09}. 

\begin{table*}
 \caption{
The best-fit parameters of the {\bf tbabs*(diskbb+powerlaw)} model of the Swift J1753.5-0127 spectra.
The value of $N_{\rm{H}}$ was fixed to 2.8 $\times$ 10${^{21}}$~cm$^{-2}$.
}
 \label{table:PL}
 \begin{flushleft}
  \begin{tabular}{cccccccc}
   \toprule

   \shortstack{MJD\\ \ \\ \ } & \shortstack{$T_{\rm{in}}$\\{\small(keV)}\\ \ } & \shortstack{$N_{\rm{disk}}$\footnotemark[$*$]\\{\small(${10^{3}}$)}} &  \shortstack{$F_{\rm{disk}}$\footnotemark[$\dagger$]\\ \ \\{\small(10$^{-9}$erg s$^{-1}$ cm${^{-2}}$)}} &  \shortstack{Observed flux\footnotemark[$\ddagger$]\\ \ \\{\small(10$^{-9}$erg s$^{-1}$ cm${^{-2}}$)}} & \shortstack{Photon\\Index\\ \ } & \shortstack{Powerlaw\\Norm\\ \ }& \shortstack{$\chi^{2}$/DOF\\ \ \\ \ }\\

   \midrule

 56037\footnotemark[$\S$] &
 0.451 ${\pm}$0.006 & 
 3.8 ${_{-0.3}^{+0.4}}$ &
 3.41 ${\pm}$0.13 & 
 2.902 ${_{-0.018}^{+0.010}}$ & 
 2.71 ${\pm}$0.04 & 
 0.70 ${\pm}$0.04  & 
 234/273\\

 56039\footnotemark[$\S$] &
 0.484 ${_{-0.009}^{+0.010}}$ & 
 2.4 ${_{-0.2}^{+0.3}}$ & 
 2.82 ${\pm}$0.11 & 
 2.884 ${_{-0.15}^{+0.008}}$ & 
 2.93 ${\pm}$0.03 & 
 0.90 ${\pm}$0.03  & 
 297/273\\

 56041\footnotemark[$\S$] &
 0.487 ${\pm}$0.009 &
 2.7 ${_{-0.2}^{+0.3}}$ &
 3.27 ${_{-0.13}^{+0.12}}$ &
 2.862 ${_{-0.03}^{+0.008}}$ & 
 2.78 ${_{-0.05}^{+0.04}}$ & 
 0.69 ${\pm}$0.04 & 
 216/273\\

 56043\footnotemark[$\S$] &
 0.448 ${_{-0.007}^{+0.008}}$ & 
 4.1 ${\pm}$0.4 & 
 3.54 ${_{-0.16}^{+0.15}}$  & 
 2.983 ${_{-0.02}^{+0.011}}$ & 
 2.67 ${\pm}$0.05 &
 0.70 ${\pm}$0.05 & 
 266/273\\

 56045\footnotemark[$\S$] &
 0.478 ${_{-0.009}^{+0.010}}$ & 
 3.4 ${\pm}$0.4 & 
 3.77 ${_{-0.19}^{+0.18}}$ & 
 2.854 ${_{-0.04}^{+0.006}}$ & 
 2.50 ${_{-0.09}^{+0.08}}$ &
 0.47 ${\pm}$0.06  & 
 281/273\\

 56047\footnotemark[$\S$] &
 0.403 ${\pm}$0.008 &
 5.4  ${\pm}$0.7 &
 3.10 ${\pm}$0.19 &
 2.708 ${_{-0.018}^{+0.010}}$ &  
 2.66 ${\pm}$0.05 &
 0.71 ${\pm}$0.05 & 
 242/273\\
 
 56049\footnotemark[$\S$] &
 0.338 ${\pm}$0.011 &
 6.7 ${\pm}$1.5 &
 1.9 ${\pm}$0.3 & 
 2.553 ${_{-0.03}^{+0.006}}$ &  
 2.61 ${\pm}$ 0.05 &
 0.90 ${\pm}$ 0.06 & 
 306/273 \\

 56050\footnotemark[$\S$] &
 0.284 ${\pm}$0.008 & 
 12.8 ${_{-1.8}^{+1.9}}$ &  
 1.8 ${_{-0.18}^{+0.19}}$ &  
 2.294 ${_{-0.03}^{+0.006}}$ &  
 2.35 ${\pm}$0.05  & 
 0.69 ${\pm}$0.05  & 
 268/273\\

 56053\footnotemark[$\S$] &
 0.201 ${\pm}$ 0.009 & 
 65 ${_{-11}^{+16}}$  &
 2.25 ${\pm}$0.16 &
 1.891 ${_{-0.04}^{+0.008}}$ &  
 2.30 ${\pm}$0.04 &
 0.57 ${\pm}$0.03  & 
 271/273\\

 56055\footnotemark[$\S$] &
 0.201 ${\pm}$ 0.008 & 
 62 ${_{-10}^{+13}}$ &
 2.16 ${_{-0.13}^{+0.12}}$ &
 1.694 ${_{-0.04}^{+0.005}}$ &  
 2.10 ${\pm}$0.04 &
 0.41 ${\pm}$0.02 &
 250/273 \\

   \midrule

 S2 \footnotemark[$\|$]&
 0.369 ${\pm{0.017}}$  &
 13 ${\pm{3}}$ &
 5.10 ${_{-0.6}^{+0.8}}$ &  
 2.95 ${_{-0.04}^{+0.02}}$ &  
 2.3 ${_{-0.4}^{+0.3}}$ &  
 0.43 ${_{-0.18}^{+0.26}}$ &  
 259/249 \\

 S3 \footnotemark[$\|$]&
 0.449 ${_{-0.03}^{+0.016}}$ &  
 4.58 ${_{-0.7}^{+1.5}}$ &  
 4.0 ${_{-0.3}^{+0.5}}$ &  
 2.52 ${_{-0.18}^{+0.03}}$ &  
 1.8 ${_{-0.5}^{+0.4}}$ &  
 0.171 ${_{-0.014}^{+0.16}}$ &  
 135/120 \\

 S4 \footnotemark[$\|$]&
 0.39 ${\pm{0.04}}$  &
 5 ${\pm{3}}$  &
 2.5 ${_{-0.9}^{+1.0}}$ &  
 2.59 ${_{-0.19}^{+0.12}}$ &  
 2.6 ${\pm{0.3}}$  & 
 0.8 ${\pm{0.3}}$ &
 173/176 \\

   \bottomrule
   \multicolumn{4}{@{}l@{}}{\hbox to 0pt{\parbox{120mm}{\footnotesize
        \par\noindent
        \footnotemark[$*$] Normalization of the {\bf diskbb} model
        \par\noindent
        \footnotemark[$\dagger$] Flux of disk in the 0.01--10~keV energy range
        \par\noindent
        \footnotemark[$\ddagger$] Flux obtained from observed spectra in the 0.6--10~keV energy range
        \par\noindent
        \footnotemark[$\S$] Observation date of Swift XRT
        \par\noindent
        \footnotemark[$\|$] Observation period of the SSC and GSC
      }\hss}}
  \end{tabular}
 \end{flushleft}
\end{table*}


\begin{table*}[hbtp]
 \caption{
The best-fit parameters of the {\bf tbabs*simpl*diskbb model} model of the Swift J1753.5-0127 spectra.
The value of $N_{\rm{H}}$ was fixed to 2.0 $\times$ 10${^{21}}$~cm$^{-2}$.
}
 \label{table:Sim}
 \begin{flushleft}
   \begin{tabular}{cccccccc}
   \toprule

   \shortstack{MJD\\ \ \\ \ } & \shortstack{$T_{\rm{in}}$\\{\small(keV)}\\ \ } & \shortstack{$N_{\rm{disk}}$\footnotemark[$*$]\\{\small(${10^{3}}$)}} & \shortstack{$F_{\rm{disk}}$\footnotemark[$\dagger$]\\ \ \\{\small(10$^{-9}$erg s$^{-1}$ cm${^{-2}}$)}} &  \shortstack{Observed flux\footnotemark[$\ddagger$]\\ \ \\{\small(10$^{-9}$erg s$^{-1}$ cm${^{-2}}$)}} & \shortstack{Photon\\Index\\ \ } & \shortstack{ $f_{\rm{SC}}$ \footnotemark[$\S$] \\ \ \\ \ }& \shortstack{$\chi^{2}$/DOF\\ \ \\ \ }\\

   \midrule

 56037\footnotemark[$\|$] &
 0.429 ${\pm}$0.008 &
 6.50 ${_{-0.4}^{+0.5}}$&
 4.73 ${\pm}$0.04 &
 2.902 ${_{-0.018}^{+0.010}}$ & 
 2.72 ${\pm}$0.17 &
 0.16 ${\pm}$0.03 &
 271/273 \\

$ $
 56039\footnotemark[$\|$] \footnotemark[$\#$] &
 0.435 ${\pm}$0.007 &
 5.9 ${_{-0.4}^{+0.5}}$ &
 4.54 ${\pm}$0.04 & 
 2.884 ${_{-0.15}^{+0.008}}$ &
 3 ${_{-0.08}^{0}}$ & 
 0.2 ${_{-0.013}^{+0.009}}$ &  
 278/234 \\

 56041\footnotemark[$\|$] &
 0.443 ${\pm 0.006}$ & 
 5.5 ${_{-0.2}^{+0.3}}$ & 
 4.57 ${\pm 0.04}$ & 
 2.862 ${_{-0.03}^{+0.008}}$ &
 3.0  ${_{-0.08}^{0}}$ &
 0.193 ${_{-0.017}^{+0.010}}$ &
 233/273 \\

$ $
 56043\footnotemark[$\|$] &
 0.425 ${_{-0.010}^{+0.009}}$&
 6.8 ${_{-0.5}^{+0.6}}$&
 4.80 ${\pm}$0.05  & 
 2.983 ${_{-0.02}^{+0.011}}$ &
 2.81 ${\pm}$0.19  &  
 0.18 ${_{-0.03}^{+0.04}}$&
 276/273 \\

$ $
 56045\footnotemark[$\|$] &
 0.478 ${_{-0.013}^{+0.012}}$  &
 4.0 ${_{-0.3}^{+0.4}}$&
 4.50 ${\pm}$ 0.05 & 
 2.854 ${_{-0.04}^{+0.006}}$ & 
 2.5 ${\pm}$ 0.3 & 
 0.12 ${_{-0.04}^{+0.05}}$ &  
 293/273 \\

$ $
 56047\footnotemark[$\|$] &
 0.389 ${_{-0.010}^{+0.009}}$&
 9.0 ${_{-0.7}^{+0.9}}$&
 4.44 ${\pm}$0.05  & 
 2.708 ${_{-0.018}^{+0.010}}$ &
 2.65 ${\pm}$0.16 & 
 0.183 ${_{-0.03}^{+0.04}}$&
 249/273 \\

 56049\footnotemark[$\|$] &
 0.308 ${_{-0.013}^{+0.012}}$& 
 19 ${\pm}$ 3 & 
 3.71 ${\pm}$0.06 & 
 2.553 ${_{-0.03}^{+0.006}}$ &
 2.67 ${\pm}$0.09 & 
 0.35 ${_{-0.04}^{+0.05}}$&
 309/273 \\

$ $
 56050\footnotemark[$\|$] &
 0.293 ${_{-0.011}^{+0.010}}$&
 18 ${_{-2}^{+3}}$&
 2.93 ${\pm}$0.05  & 
 2.294 ${_{-0.03}^{+0.006}}$ &
 2.32 ${_{-0.07}^{+0.06}}$&
 0.33 ${\pm}$ 0.03 &  
 275/273 \\

$ $
 56053\footnotemark[$\|$] &
 0.219 ${_{-0.011}^{+0.010}}$&
 53 ${_{-9}^{+13}}$&
 2.63 ${\pm}$ 0.06 & 
 1.891 ${_{-0.04}^{+0.008}}$ &
 2.26 ${\pm}$ 0.05 & 
 0.33 ${_{-0.02}^{+0.03}}$ &  
 272/273 \\
 
 56055\footnotemark[$\|$] &
 0.221 ${\pm}$0.009& 
 43 ${_{-7}^{+9}}$& 
 2.19 ${\pm}$0.05& 
 1.694 ${_{-0.04}^{+0.005}}$ &
 2.06 ${\pm}$0.05 &
 0.293 ${_{-0.018}^{+0.019}}$&
 260/273 \\

   \midrule

   S2 \footnotemark[$\#\#$]&
   0.38 ${\pm}$0.03 &
   12  ${_{-3}^{+4}}$& 
   5.3 ${\pm}$0.3 &
   2.95 ${_{-0.04}^{+0.02}}$ &
   2.3  ${\pm}$0.5 &
   0.10 ${_{-0.04}^{+0.07}}$ &
   262/249 \\

   S3 \footnotemark[$\#\#$]&
   0.45 ${_{-0.03}^{+0.02}}$& 
   4.6 ${_{-0.8}^{+1.6}}$& 
   4.2 ${_{-0.2}^{+0.3}}$& 
   2.52 ${_{-0.18}^{+0.03}}$ &
   1.8 ${_{-0.7}^{+0.4}}$& 
   0.074 ${_{-0.009}^{+0.04}}$ &
   136/120 \\

   S4 \footnotemark[$\#\#$]&
   0.35 ${_{-0.06}^{+0.05}}$& 
   12  ${_{-5}^{+12}}$& 
   3.9 ${\pm}$0.3 & 
   2.59 ${_{-0.19}^{+0.12}}$ &
   2.7 ${\pm}$0.4 & 
   0.30 ${_{-0.06}^{+0.05}}$ &
   173/176 \\

   \bottomrule
   \multicolumn{4}{@{}l@{}}{\hbox to 0pt{\parbox{120mm}{\footnotesize
       \par\noindent
        \footnotemark[$*$] Normalization of the {\bf diskbb} model
        \par\noindent
        \footnotemark[$\dagger$] Flux of disk in the 0.01--10~keV energy range 
        \par\noindent
        \footnotemark[$\ddagger$] Flux obtained from observed spectra in the 0.6--10~keV energy range
        \par\noindent
        \footnotemark[$\S$] Scattered fraction
        \par\noindent
        \footnotemark[$\|$] Observation date of Swift XRT
        \par\noindent
        \footnotemark[$\#$] The fitting range from 1.0 keV to 9.4 keV
        \par\noindent
        \footnotemark[$\#\#$] Observation period of the SSC and GSC
      }\hss}}

  \end{tabular}
 \end{flushleft}
\end{table*}

\begin{figure}
   \begin{center}
      \FigureFile(90mm,60mm){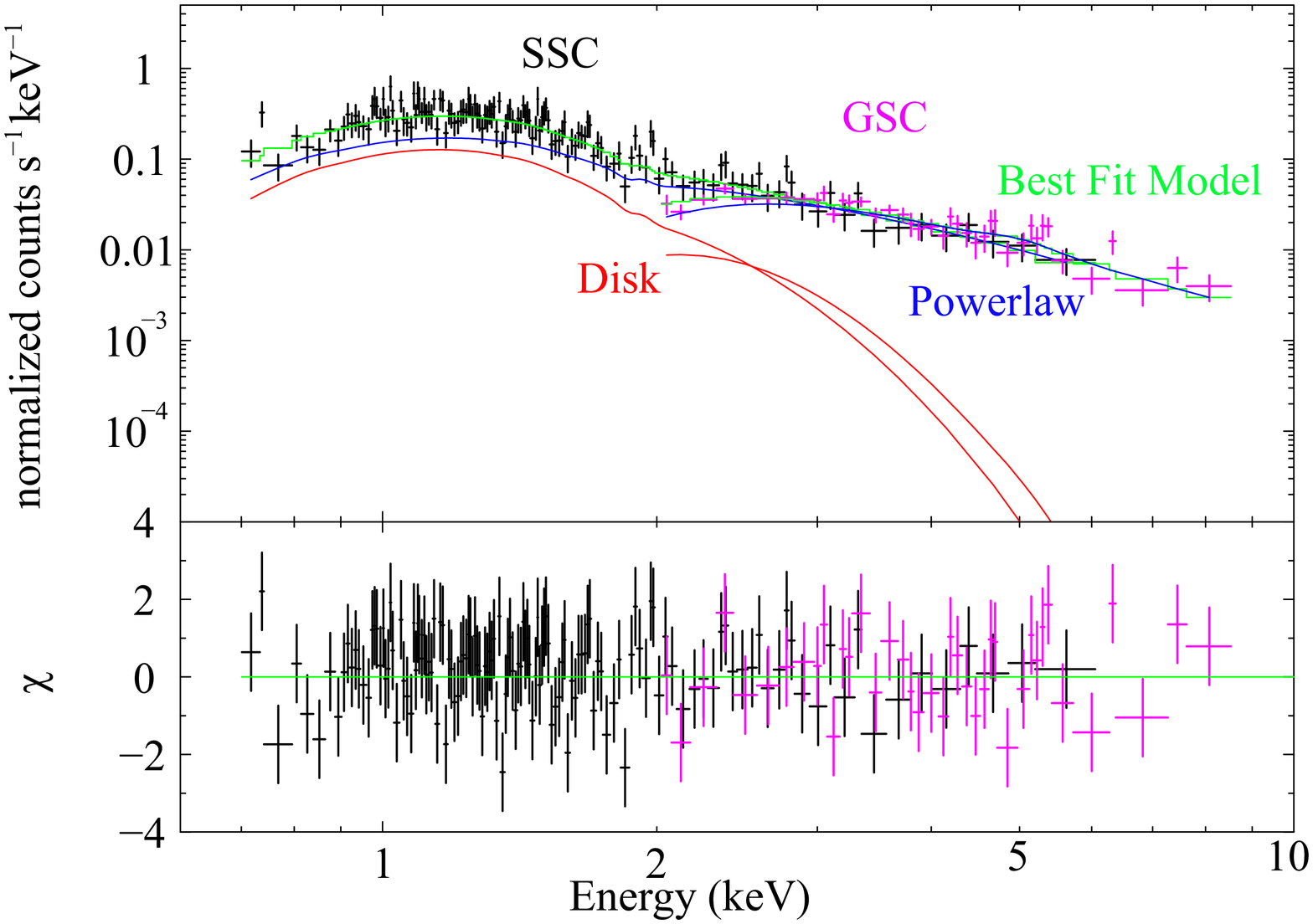}
   \end{center}
   \caption{
	The spectrum observed by the SSC (black) and GSC (magenta) in S4.
	The best-fit model of {\bf tbabs*(diskbb+powerlaw)}, the {\bf tbabs*diskbb} component, and the {\bf powerlaw} component are shown as green, red, and blue lines.
	Residuals between the data and best-fit model are shown in the lower panel.
   }\label{FIG:MAXICntsSpe}
\end{figure}

To obtain the best-fit parameters in S2 and S3 that were not observed by the XRT,
we analyzed the spectra observed by the SSC and GSC in S2 and S3 in the same way for S4.
The S1 spectrum in the energy range of 0.7--10.0 keV could not be obtained, because the SSC data lacked in S1.
First, to check if $T_{\rm{in}}$ and $F_{\rm{disk}}$ obtained from the SSC and GSC spectra are 
consistent with those from the XRT spectra,
we fitted the {\bf tbabs*(diskbb+powerlaw)} and {\bf tbabs*simpl*diskbb} models to the spectra observed by the SSC and GSC in S4.
Figure \ref{FIG:MAXICntsSpe} shows the time average spectrum observed.
The best-fit parameters of the {\bf tbabs*(diskbb+powerlaw)} and {\bf tbabs*simpl*diskbb} models are listed in tables \ref{table:PL} and \ref{table:Sim}, respectively.
The difference between the observational results of the SSC and GSC and those of the XRT in S4 was less than 20$\%$.
The values of ${T_{\rm{in}}}$ and $F_{\rm{disk}}$ in S2 and S3, which are almost the same as those in S4 within error bars, are included in tables \ref{table:PL} and \ref{table:Sim}.

\begin{figure}
   \begin{center}
      \FigureFile(90mm,60mm){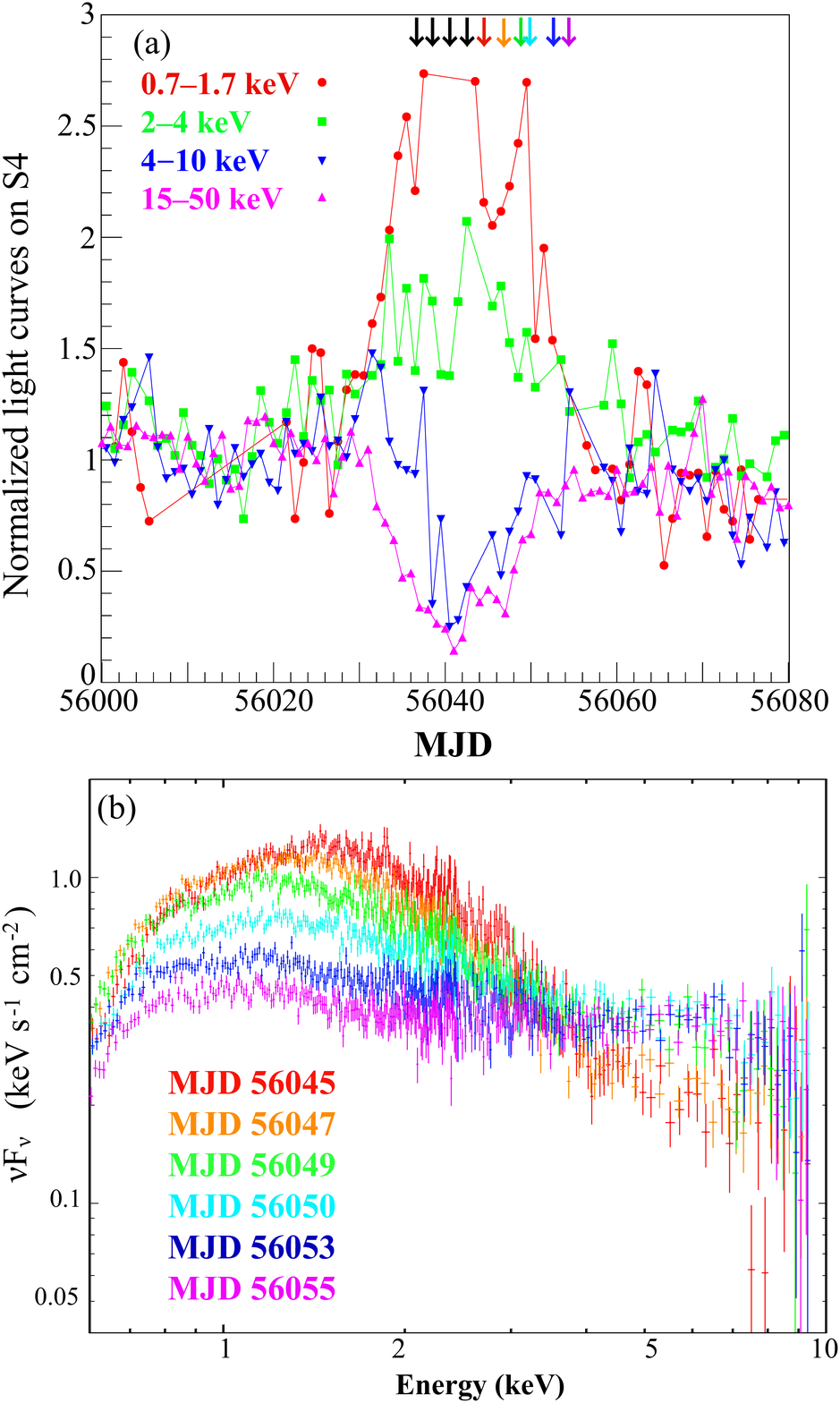}
   \end{center}
   \caption{
      (a) The light curves in energy bands of 0.7--1.7~keV (red), 2--4~keV (green), 4--10~keV (blue) and 15--50~keV(magenta) normalized by the average value excluding the spectral softening. 
      Arrows indicate the XRT observations on MJD 56045 (red), 56047 (orange), 56049 (green), 56050 (cyan), 56053 (blue), 56055 (magenta) and the other dates (black).
      These colors correspond to the spectrum in figure (b).
      (b) The Swift XRT spectra on MJD 56045 (red), 56047 (orange), 56049 (green), 56050 (cyan), 56053 (blue) and 56055 (magenta).
   }\label{FIG:EEUFCH}
\end{figure}

To investigate the differences of timescale between from the hard to soft and from the soft to hard,
we overlaid the light curves of the various energy ranges in S4 normalized to the mean count rates, 
where the mean count rates are calculated by taking an average over $\sim$40~days (MJD 56000--56020 and 56060--56080).
The result can be seen in figure \ref{FIG:EEUFCH}a.
The hump in the 0.7--4~keV band and the dip in the 15--50~keV band correspond to the growth and decline of the {\bf diskbb} emission and the Comptonization emission, respectively.
In the 4--10~keV energy range, the {\bf diskbb} emission accounted for 10$\%$ of the total intensity and the rest was the Comptonized emission.
The errors of the light curves in the 0.7--10~keV and 15--50~keV bands are $\sim$30$\%$ and $\sim$10$\%$, respectively.
These include the statistical and systematic errors of the incoming photons and non X-ray background events.
Since the light curves in the 0.7--1.7~keV and 2--4~keV bands are symmetric with respect to the peak on MJD 56040,
the timescales of the growth and decline of the {\bf diskbb} component are similar.
The light curve in the 15--50~keV band seems symmetric, while the light curve in the 4--10~keV band seems asymmetric.
The intensity in the 4--10~keV band dropped off around MJD 56038, and the intensity increased after MJD 56040 as did that in the 15--50~keV band.
There might be hysteresis between the state transitions from the low/hard to the high/soft state and from the high/soft to the low/hard state.

\section{Discussion}
\subsection{Summary of the observational results \label{SEC:SumObs}}
Through the analyses described in subsections (\ref{SEC:MS} and \ref{SEC:3}), the transitions from the low/hard state to the likely high/soft state in the short-term softenings were quantified in the following manner.

\begin{enumerate}
\item The short-term softening is characterized by an intensity dip of hard X-rays (15--50~keV) and also an intensity hump of soft X-ray (0.7--4~keV) for a few tens of days (subsection \ref{SEC:MS}).
\item The light curves and HID show that the short-term softenings occurred four times between 2009 and 2013 (subsection \ref{SEC:MS}).
\item The fitting results of the X-ray spectra, which were reproduced well with a model consisting of optically thick disk emission and its Comptonized component, were explained by a gradual decrease of the {\bf diskbb} emission (subsection \ref{SEC:3}).
\end{enumerate}

From these results, the short-term softenings are explained not by eclipse (\cite{SH13}) but rather by the state transitions associated with an increase of the mass accretion rate over a short-term timescale (a few tens of days).

\subsection{Interpretation on the fitting results \label{SEC:ISCO}}
\begin{figure}
   \begin{center}
      \FigureFile(90mm,60mm){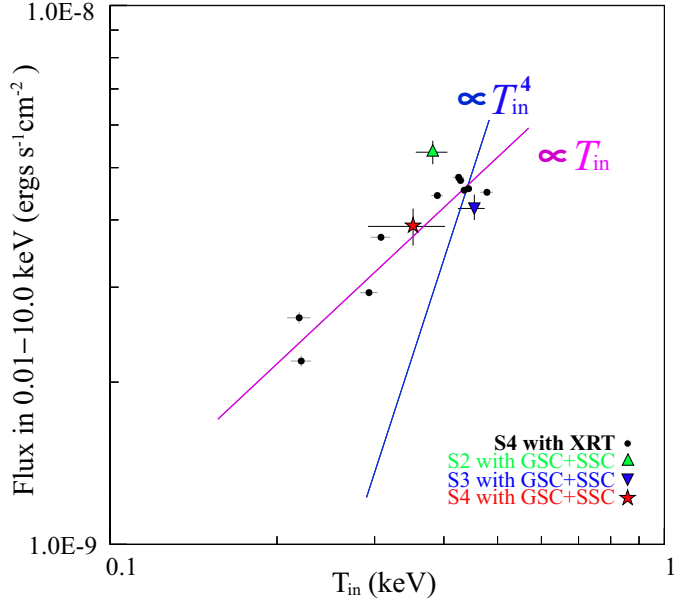}
   \end{center}
   \caption{
    Relation between $F_{\rm{disk}}$ in the 0.01--10~keV energy range and $T_{\rm{in}}$ using the best-fit parameters for the short-term spectral softenings.
    Black points show the fitting results of S4 observed by the XRT.
    Green, blue, and red points show the fitting results for observations by the SSC and GSC in S2, S3, and S4, respectively.
    Magenta and blue lines show the powerlaw functions with an index of one and four, respectively.
   }\label{FIG:FT}
\end{figure}

Figure \ref{FIG:FT} shows the relation between $F_{\rm{disk}}$ and $T{_{\rm{in}}}$ of Swift J1753.5-0127 during the short-term softening.
The data were fitted by $F_{\rm{disk}}$ $\propto$ $T{_{\rm{in}}}$ $^{ 0.97 \pm 0.02}$.
Note that a $F_{\rm{disk}}$ $\sim$ $T{_{\rm{in}}}$ $^{4/3}$ relation has also been reported 
from XTE J1817-330, possibly as the result of a constant mass accretion rate \citep{Ca09}.
The lines corresponding to the powerlaw indices of 1 and 4 are overlaid on the figure \ref{FIG:FT}.
The accretion disk extending down to the ISCO in the low/hard state has been suggested \citep{Re09}.
If $R_{\rm{in}}$ reaches at the ISCO, then $F_{\rm{disk}}$ $\propto$ $T_{\rm{in}}$ $^4$ is expected.
However, the data during the short-term softening seems to follow the curve of $F_{\rm{disk}}$ $\propto$ $T_{\rm{in}}$, 
which may suggested the truncate disk.
The value of $f_{\rm{SC}}$ increased from 0.12 to 0.30 as $R_{\rm {in}}$ increased, suggesting that the covering fraction of the corona increases.
These results imply that the mass accretion rate is insufficient to stabilize the inner edge of the disk at the ISCO,
 and there is a possibility that $R_{\rm {in}}$ is larger than the ISCO.

The high/soft state may be realized between MJD 56037 and 56047, in which the softest spectra were obtained and N${_{\rm{disk}}}$ appeared to plateau in figure \ref{FIG:LCPa}.
By using the value of N${_{\rm{disk}}}$ in the plateau, MJD 56037--56047, 
$r_{\rm{in}}$ is estimated as $R_{\rm{in}}$ $\sim$ 70$\pm3$ $\times$ ($D$ / 10~kpc) $\times$ cos$^{-0.5}$ $\theta$~km, where $D$ is the distance to the source, and $\theta$ is the inclination angle.
The true inner radius ($r_{\rm{in}}$) can be calculated as 1.18 $R_{\rm{in}}$ assuming the friction free boundary condition \citep{Ku98} and a color harding factor of 1.7 (\cite{SHI95}).
Assuming that $D$ is 8.5 kpc \citep{Rey10} and $\theta$ is 45$^{\circ}$, we obtained $r_{\rm{in}}$ $\sim$82~km.
When the black-hole mass is 10 $\MO$ \citep{Rey10} and the black-hole has no spin, $r_{\rm{in}}$ would be close to the ISCO; although this is not yet conclusive due to the large uncertainties of the mass, $\theta$ and $D$.

\section{Summary} 
We have presented a detailed spectral and timing analysis of the SSC, GSC, and BAT monitoring observations during MJD 55500--56400 and the 10 XRT pointing observations around MJD 56040.
The light curves show that state transitions of Swift J1753.5-0127 in the short-term softening occurred four times between 2009 and 2013.
We performed a spectral analysis of a the short-term softening, and found that softening events can be explained not by eclipse (\cite{SH13}) but by state transitions.
Although the parameters during the softening events are close to those of the high/soft state, the results imply that the inner disk radius is larger than the ISCO.
Since there are large uncertainties in the mass, $\theta$, $D$ and $N_{\rm{H}}$, 
the possibility that the inner disk radius reaches to the ISCO can not be completely ruled out.
Therefore, the softening events are triggered by a short-term increase of 
the mass accretion rate which is insufficient to create the standard accretion disk.

This research has made use of MAXI data provided by RIKEN, JAXA and the MAXI team, and use of Swift data provided by NASA, ASI, and the University of Leicester. 
This research was partially supported by the Japan Society for the Promotion of Science.


\end{document}